\documentclass{emulateapj}
\usepackage{graphicx,color}
\usepackage{amssymb,amsmath}
\usepackage[colorlinks,hyperfootnotes=false,citecolor=blue,linkcolor=blue]{hyperref}

\begin{document}
\shorttitle{Strong Lensing Analysis of A1722}
\shortauthors{Zitrin et al.}

\slugcomment{Submitted to the Astrophysical Journal}

\title{Strong Lensing Analysis of the Galaxy Cluster MACS J1319.9+7003 and the Discovery of a Shell Galaxy}

\author{Adi Zitrin\altaffilmark{1,2}}
\altaffiltext{1}{Cahill Center for Astronomy and Astrophysics, California Institute of Technology, MC 249-17, Pasadena, CA 91125, USA; adizitrin@gmail.com}
\altaffiltext{2}{Hubble Fellow}


\begin{abstract}
We present a strong-lensing (SL) analysis of the galaxy cluster MACS J1319.9+7003 ($z=0.33$, also known as Abell 1722), as part of our ongoing effort to analyze massive clusters with archival {\it HST} imaging. We spectroscopically measured with Keck/MOSFIRE two galaxies multiply-imaged by the cluster. Our analysis reveals a modest lens, with an effective Einstein radius of $\theta_{e}(z=2)=12\pm1\arcsec$, enclosing $2.1\pm0.3\times10^{13}$ $M_{\odot}$. We briefly discuss the SL properties of the cluster, using two different modeling techniques, and make the mass models publicly-available\footnote{ftp://wise-ftp.tau.ac.il/pub/adiz/MACS1319/}. Independently, we identified a noteworthy, young Shell Galaxy (SG) system forming around two likely interacting cluster members, 20$\arcsec$ north of the BCG. SGs are rare in galaxy clusters, and indeed, a simple estimate yields that they are only expected in roughly one in several dozen, to several hundred, massive galaxy clusters (the estimate can easily change by an order-of-magnitude within a reasonable range of characteristic values relevant for the calculation). Taking advantage of our lens model best-fit, mass-to-light scaling relation for cluster members, we infer that the total mass of the SG system is $\sim1.3\times10^{11}$ M$_{\odot}$, with a host-to-companion mass ratio of about 10:1. Despite being rare in high density environments, the SG constitutes an example to how stars of cluster galaxies are being efficiently redistributed to the Intra Cluster Medium. Dedicated numerical simulations for the observed shell configuration, perhaps aided by the mass model, might cast interesting insight on the interaction history and properties of the two galaxies. An archival HST search in galaxy cluster images can reveal more such systems. \vspace{0.05cm}
\end{abstract}


\section{Introduction}\label{intro}
Shell Galaxies (SGs) are, typically, elliptical galaxies, surrounded by low surface brightness shells, or at times, cones, seen as concentric arcs around the SG center. The first SGs were noted some 50-60 years ago \citep[][see also \citealt{Zwicky1956UmbrellaShell}]{Arp1966Shell}, and then in earnest around the early 1980's \citep{MalinCarter1980NaturShells}, and have been then studied observationally, analytically, and by numerical simulations \citep[e.g.][and references therein]{MalinCarter1983ShellCatalog,Quinn1984Shell,SchweizerFordShells,AthanassoulaBosma1985ShellReview,DuprazCombes1986Shells,HernquistQuinn1988ShellsSpheric}. Significantly improved computer power in recent years has become particularly useful for simulating such galaxies with greater detail \citep[e.g.][]{Cooper2011ShellSimNGC7600Like,Ebrova2012LineProfilesShells,Ebrova2013Thesis}, generating renewed interest in these systems \citep[see also][]{Canalizo2007Shells,Sikkema2007Shells,Bennert2008Shells,Foster2014UmbrellaShell}. 

The shells are a particular tidal feature that forms as a result of an interaction between two galaxies \citep[see for a recent review of shell galaxies][and references therein]{Ebrova2013Thesis}, in particular a highly-radial, minor merger \citep[][but see also \citealt{HernquistSpergel1992}]{Quinn1984Shell}. The shells consist of stars stripped by the interaction, oscillating in the system's potential well and forming faint envelopes near the turnaround radii \citep[e.g.][]{DuprazCombes1986Shells,HernquistQuinn1988ShellsSpheric}. Shells are relatively common around elliptical galaxies (at least $10\%$ show shells, e.g. \citealt{MalinCarter1983ShellCatalog,AthanassoulaBosma1985ShellReview,Ebrova2013Thesis}), but are quite rare around spiral or disk galaxies \citep[cf.][]{SchweizerSeitzer1988,Foster2014UmbrellaShell,Fardal2007Shell}. Despite being seen mostly around elliptical galaxies, most shells have been observed in the field rather than in clusters of galaxies \citep[e.g.][]{MalinCarter1983ShellCatalog,AthanassoulaBosma1985ShellReview}. This is likely a result of various factors, primarily the low cross-section for small impact parameter galaxy encounters within the cluster ($\pi r_{core}^2$, where $r_{core}$ is the typical galaxy's core size), the collisionlessness of dark matter and stars, combined with high encounter velocities which lower the chances for merger within the cluster. In addition, it is conceivable the Intra Cluster Light may also play a role in smoothing the shell structure in clusters so it becomes harder to observe due to lack of contrast. 

The number of shells, and distance between them can shed light on the interaction or merger history of the two galaxies as in each passage of the smaller galaxy at the host's center \citep{Gu2013ComaShell}, more material is stripped to form an expanding front \citep[e.g.][]{Quinn1984Shell,Ebrova2013Thesis}. The shape of the shell, especially in the case of narrow cones, adds useful information that can be then used to tighten the constraints on the initial configuration, relative masses, and velocities \citep{HernquistQuinn1988ShellsSpheric,Ebrova2012LineProfilesShells}, although significant degeneracies exist. Also, color information and gradients, if seen, might add information relevant for a population synthesis of the shell stars and the system's history \citep[e.g.][]{Bilek2016shellcore,Sikkema2007Shells}.

Here, we present a SG system caught relatively-early on, so that only one highly symmetric shell is seen on each side of the system where the distance of the shell on one side is half the distance on the other side, and the two interacting galaxies are both still observed \citep{Bilek2016shellcore}. The SG is formed in a massive galaxy cluster, MACS J1319.9+7003 (hereafter M1319, $z=0.33$; \citealt{Mantz2010,Ebeling2010FinalMACS}; also known as Abell 1722 \citealt{Abell1989cat}), where, as mentioned, SGs are generally considered less common. 

The system was identified in the framework of our ongoing effort \citep[e.g.][]{Zitrin2016MACS2135} to lens-model massive clusters with available \emph{Hubble} imaging, towards the launch on the James Webb Space Telescope (JWST). Since one of the main goals of JWST is to target galaxies in the era of reionization, strong lensing (SL) by galaxy clusters will continue being of increasing importance for detecting the faintest, highest-redshift galaxies. In addition, M1319 has another interesting aspect due to its high ecliptic latitude, where both the IR background is $\sim1.3$ magnitudes fainter \citep{Windhorst2011WFC3IR} and dust extinction from our Galaxy is minimized, so it may become a good candidate for high-redshift searches in the future. Here we map the projected mass distribution of M1319, from measurements of two strongly lensed galaxies we identify and measure below. The model is made publicly available, for example, for future studies of matter distributions in clusters \citep{Donahue2014CLASHX,Umetsu2016CLASH,Meneghetti2014CLASHsim}, lensed background sources or lensing-efficiency measurements \citep{Coe2014FF,Lotz2016HFF}, but also, if so desired, it can then be used to numerically simulate the details and environment of the SG with greater detail. 

The presented SG, although rare, supplies an interesting example of how galaxies can merge, evolve, and lose their material to the ICM \citep[e.g][]{Edwards2016BCGPopulation}, warranting further study \citep[see also][]{Gu2013ComaShell}. However, despite the interesting case of the SG, note that we will only present and give its basic characteristics, and the paper mainly concentrates on the SL properties of the cluster, so that we leave a detailed examination (and possibly, numerical simulation) of the SG system to other dedicated work.

The paper is organized as follows: In \S \ref{obs} we review the different observations, and show the spectra of the identified multiply-imaged galaxies. In \S \ref{lensmodel} we construct a mass model for the cluster and summarize its properties. In \S \ref{discussion} we summarize and discuss the results, including an estimate of the occurrence of SGs in galaxy clusters. Throughout we use a standard $\Lambda$CDM cosmology with $\Omega_{\rm m0}=0.3$, $\Omega_{\Lambda 0}=0.7$, $H_{0}=100$ $h$ km s$^{-1}$Mpc$^{-1}$, $h=0.7$, and magnitudes are given using the AB convention. 1\arcsec\ equals 4.75 kpc at the redshift of the cluster, $z_{l}=0.33$. Unless noted otherwise, errors are $1\sigma$.

\begin{figure*}
 \begin{center}
   \includegraphics[width=177mm,trim=0cm 0cm 0cm 0cm,clip]{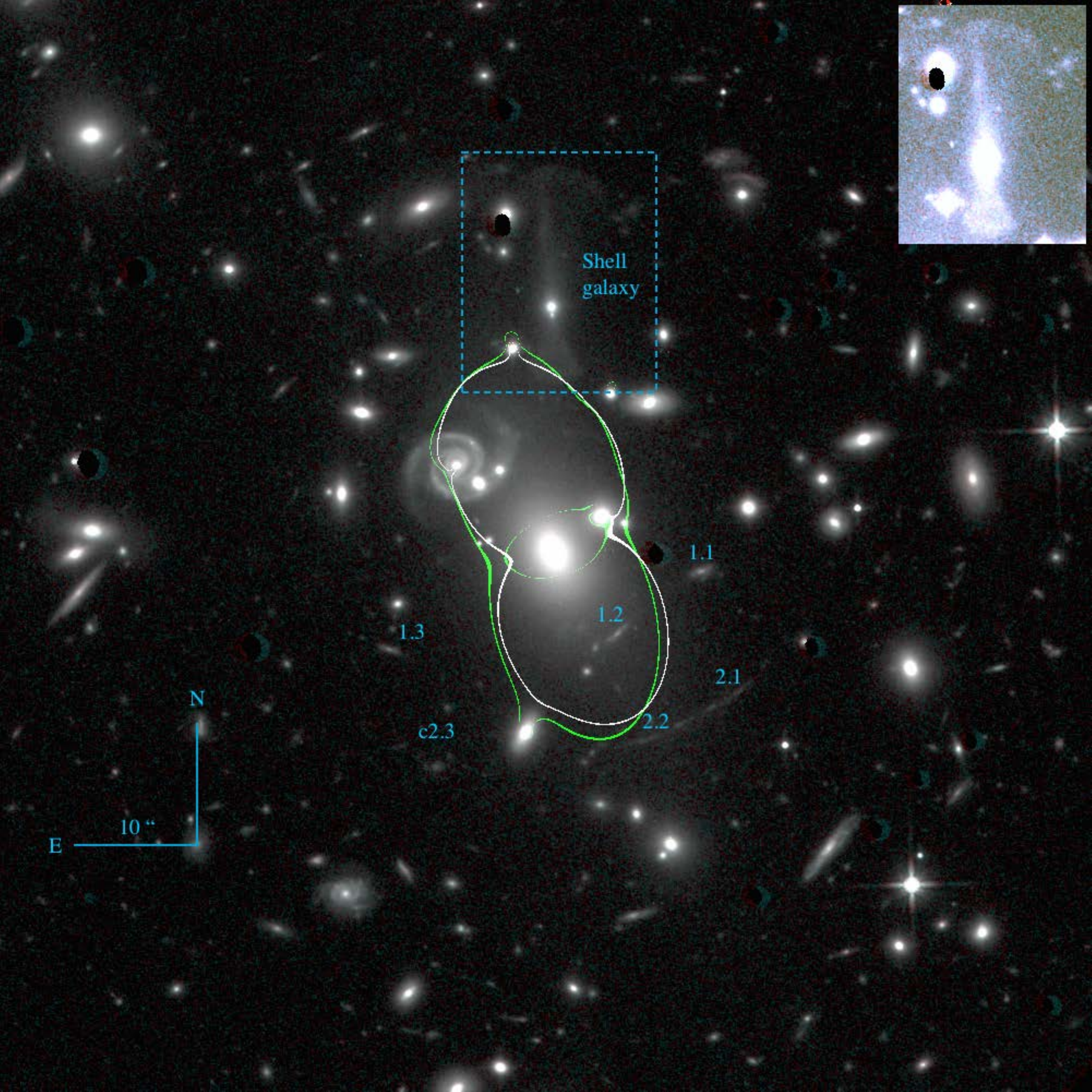}
 \end{center}
\caption{Central field of the galaxy cluster M1319. The SG is marked with a dashed rectangle whose length is $\simeq20\arcsec$ (1\arcsec\ is 4.75 kpc at the cluster's redshift), and is inset for show also as a stamp in the upper-right corner, with higher contrast. The image also shows two sets of multiply-imaged galaxies we identified and measured spectroscopically with Keck/MOSFIRE. We constructed two complementary SL models (see text for details) for the cluster using those two system, excluding image 2.3 (rendered a candidate, less secure identification marked with ``c'' above). The critical curves from the models are marked in white and green, for a source at redshift $z=1.55$ (system 1), enclosing an area with an effective Einstein radius of $\theta_{e}(z=1.55)=11\pm1\arcsec$. The image is constructed from F110W and F140W HST/WFC3 imaging (see \S \ref{obs}).}\vspace{0.05cm}
\label{fig1}
\end{figure*}

\section{Data and Observations}\label{obs}

We primarily use archival \emph{Hubble Space Telescope} (HST) observations of the galaxy cluster M1319, in which we identified the lensed features and noted the SG. These data include imaging in four bands from HST programs 10266 and 10491 (PI: Ebeling) available through the Hubble Legacy Archive: a F606W image (total exposure time 1200s), taken on 2005-11-04, and a F814W image (total exposure time 1440s), taken on 2011-01-22, with the ACS/WFC; and F110W and F140W, 705.88s each, taken on 2011-07-17 with the WFC3/IR. 

We ran SExtractor \citep{BertinArnouts1996Sextractor} in dual-mode to obtain the photometry of objects in the cluster field, useful for identifying multiply-imaged galaxies as well as the red-sequence cluster members (\S  \ref{lensmodel}). We then use the resulting catalogs as input and run the Bayesian Photometric Redshift program (BPZ; ; \citealt{Benitez2000,Coe2006}), to derive photometric redshifts, especially examining multiple-image candidates.

We observed the cluster field with the Multi-Object Spectrometer For Infra-Red Exploration (MOSFIRE; \citealt{KeckMOSFIRERef}) on the Keck 1 telescope, for approximately half an hour, consisting of sets of 120s exposures, on 2015 June 10, placing a slit along the SG system, and on multiple-images 1.1 and 2.2 seen in Figure \ref{fig1}. Observations were carried out in the H-band, primarily to examine if a prominent Paschen-beta (Pa$\beta$) line was present in the SG and to capture redshifted optical or long-UV spectral lines from the multiply-imaged systems. We adopted a dither pattern of $\pm2$\arcsec\ along the slit. 

\begin{figure*}
 \centering
   \includegraphics[width=140mm,trim=3cm 3cm 2cm 2cm,clip]{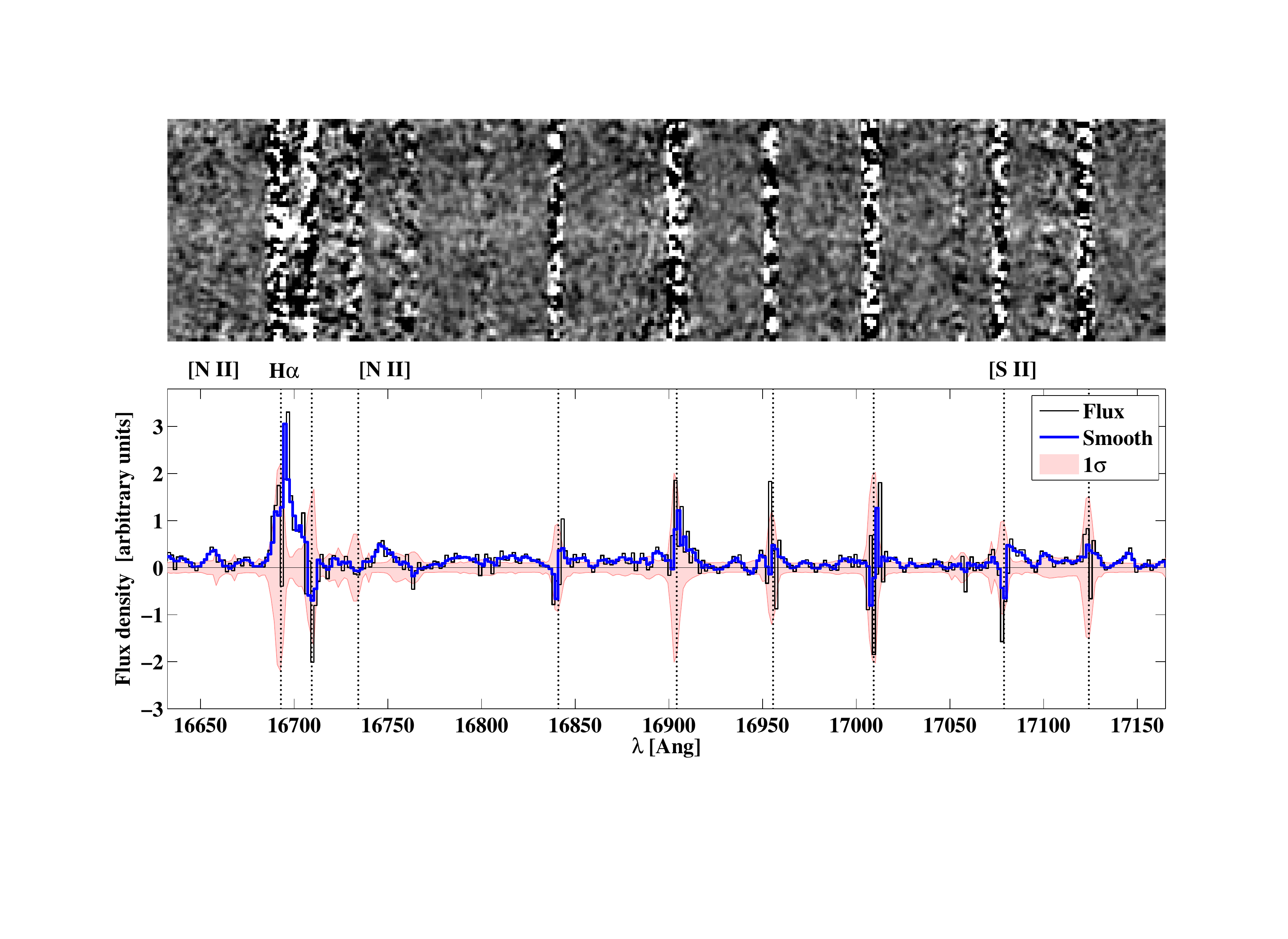}
    \includegraphics[width=140mm,trim=2.5cm 3cm 2cm 2cm,clip]{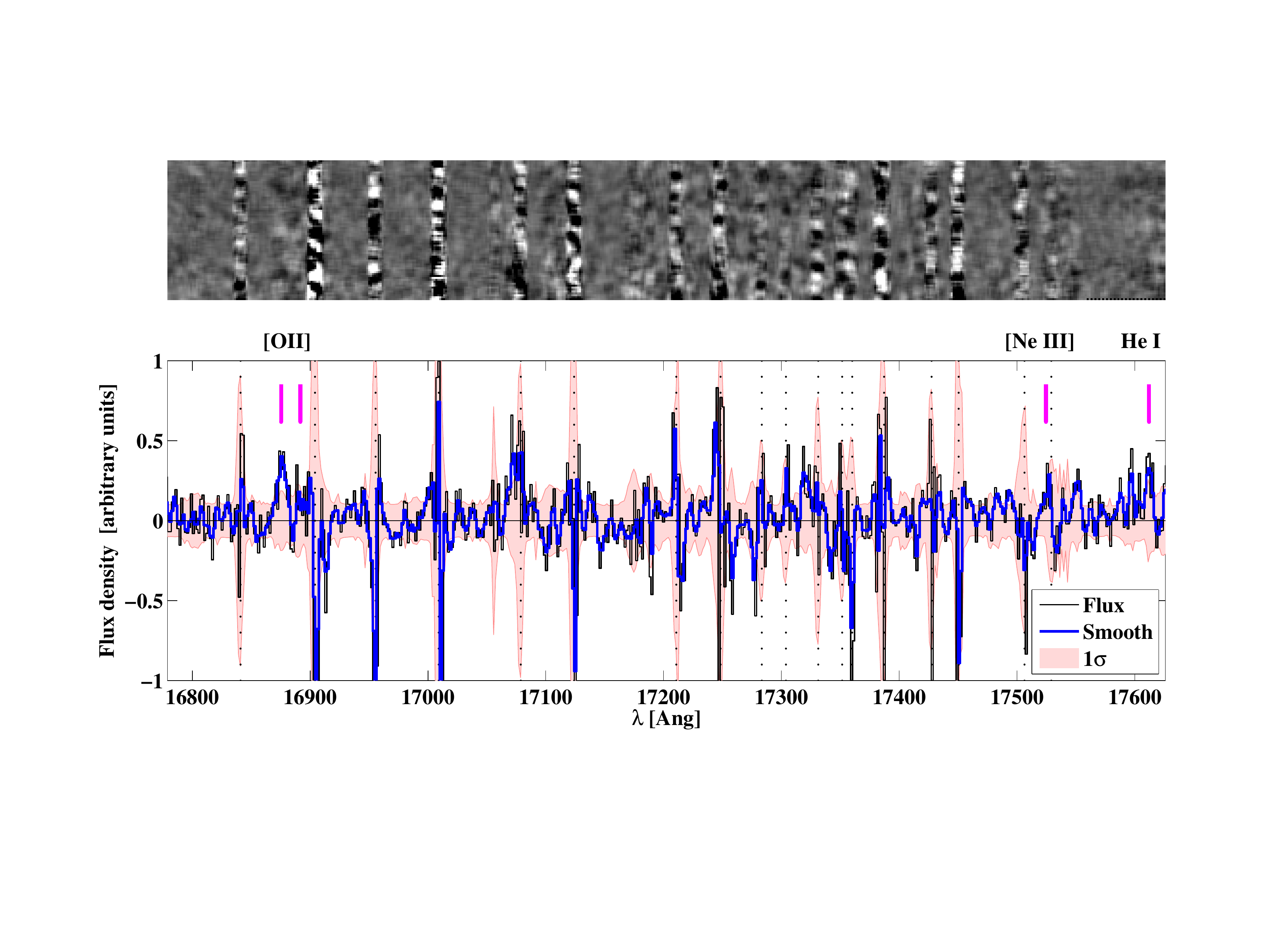}
\caption{Spectra of multiple images 1.1 (upper subfigure) and 2.2 (bottom subfigure). Each subfigure shows both the 2D (upper inset) and 1D spectra (bottom inset; black curve), including a slightly smoothed version of the 1D spectra for illustrative purposes (blue curve). The 1$\sigma$ error is also shown as a pink shaded region. In image 1.1 we identify the two [N II] doublet lines ($\lambda\lambda$6549,6583 \AA) bracketing the prominent H$\alpha$ line ($\lambda$6563 \AA), and additionally, the [S II] ($\lambda$6717 \AA) doublet-line seems to be present as well (the expected position of the other doublet line, [S II] ($\lambda$6731 \AA), falls on a skyline).  These correspond to a redshift of $z=1.55$ in excellent agreement with the photometric redshift (and 95\% C.L.)  of 1.44 [1.20-1.68]. Image 2.2 is fainter and line identification is less secure. We utilize the redshift prediction from our lens model, $z\sim3.4-3.6$, to best-fit a redshift of $3.52$ following the likely -- but tentative -- identification of the [O II] doublet ($\lambda\lambda$3726,3729 \AA), the He I ($\lambda$3889 \AA), and [Ne III] ($\lambda$3868 \AA) lines ([Ne IV] and [Ne V] are also covered in the slit but are not identified). We add purple markers to note the position of these faint lines.}
\label{fig2}
\end{figure*}

Data reduction was performed using the official MOSFIRE  pipeline\footnote{http://www2.keck.hawaii.edu/inst/mosfire/drp.html}. For each flat-fielded slit we extracted the 1D spectrum using a 11 pixel boxcar ($\simeq1\arcsec$) centered on the target, and a similar procedure was adopted in quadrature to derive the 1$\sigma$ error distribution. We use two stars with known magnitudes, on which slits were placed in order to track possible drifts, for estimating the absolute depth of our observations. We reach a $3\sigma$ flux density limit of $\sim2.1\times10^{-18}$ erg cm$^{-2}$ s$^{-1}$ \AA$^{-1}$ between skylines, which for a marginally-resolved line FWHM=5\AA\ line translates into a 3$\sigma$ line flux limit of $\sim1.8\times10^{-18}$ erg cm$^{-2}$ s$^{-1}$, in good agreement with the MOSFIRE exposure time calculator (yielding 3$\sigma$ $\sim2\times10^{-18}$ erg cm$^{-2}$ s$^{-1}$) and with our expectations based on previous observations and taking into account the different exposure times (e.g. \citealt{Zitrin2015CIII}). The absolute calibration also agrees to within 10\% typically, with the nominal MOSFIRE absolute calibration files (C. Steidel, private communication). No prominent lines were detected in the SG slit, disfavoring exotic, AGN-related mechanisms for the observed cones, such as ionization cones or jet-related features (a typical FWHM of an AGN can reach often order 1000-3000 km/s). This spectrum is thus not shown. The reduced 2D and 1D spectra of multiple images 1.1 and 2.2 are shown in Fig. \ref{fig2}, corresponding to $z_{s}=1.55$ for system 1, and $z_{s}=3.52$ for system 2, although the latter is less certain, as we show and discuss in Fig. \ref{fig2}.

\vspace{0.05cm}

\section{Lens Model} \label{lensmodel}
To construct a SL model for M1319 we use primarily the light-traces-mass (LTM) approach by \citet[][see also \citealt{Broadhurst2005a,Zitrin2014CLASH25}]{Zitrin2009_cl0024}. Full details can be found in these papers. Here we describe the method with brevity. 

We start with the cluster galaxies, chosen by following the red-sequence in a color-magnitude diagram. Each member galaxy is parametrized as power-law mass density distribution, with a weight in proportion to its luminosity, and the superposition of all galaxies makes the total galaxy component of the model. The power-law exponent is the same for all galaxies and is a free parameter of the model. This mass density map is then smoothed with a 2D Gaussian, whose width is also a free parameter of the model, to obtain the smooth dark matter component (this is why this method is referred to as LTM - both the galaxy and dark matter component follow the light). The two components are then combined with a relative weight -- the third free parameter of the method, which along with the overall normalization, brings the number of free parameters to four. A two component external shear is usually also added to allow for further flexibility, and to improve the fit we sometimes allow single bright galaxies to be freely weighted in the minimization and deviate from the nominal mass-to-light ratio adopted (in our case, only the BCG is left to be freely weighted). We also leave the ellipticity of the BCG a free parameter.

We first ran a model fixing the redshift of system 1 to $z=1.55$ as indicated by our MOSFIRE data (Fig. \ref{fig2}), but allowing the redshift of system 2 to vary given the line identification in this system was ambiguous. We ran various models with different priors and found that they place system 2 at $z\sim3.5$, a noticeably higher redshift than initially implied by its $z\sim1.6$ [1.2--1.9] photometric redshift. Following the model's preference we searched more carefully for spectroscopic solutions around $z\sim3.5$ for system 2, and managed to identify the [O II] doublet and other faint lines, seen in the expected position. We thus infer -- even if somewhat more tentatively -- a redshift of $z=3.52$ for system 2. We fixed its redshift to this value and reran the model whose resulting critical curves are seen in Figure \ref{fig1}. The minimization of the model included about a couple thousand Monte Carlo Markov Chain (MCMC) steps, and the final LTM model has an image reproduction \emph{rms} of 0.6\arcsec. 

We also construct a complementary, fully-parametric model using our so-called PIENDeNFW pipeline \citep[see][]{Zitrin2014CLASH25}: Pseudo Isothermal Elliptical Mass Distributions are used to model the cluster galaxies, scaled by their light \citep[following the prescription of][]{Jullo2007Lenstool}, and the DM component is an analytic elliptical NFW \citep{Navarro1996} form. This method is particularly relevant for our case since it adopts well-tested scaling relations  \citep[see also][]{Monna2016} for the cluster galaxies and thus gives an empirical separation between the galaxies, and cluster-scale dark components, so that we can estimate directly what is the mass of the SG. The final \emph{rms} for this model is $\sim1\arcsec$, slightly higher than that of the LTM model.

\begin{figure}
 \centering
   \includegraphics[width=82mm,trim=2cm 1cm 2cm 2cm,clip]{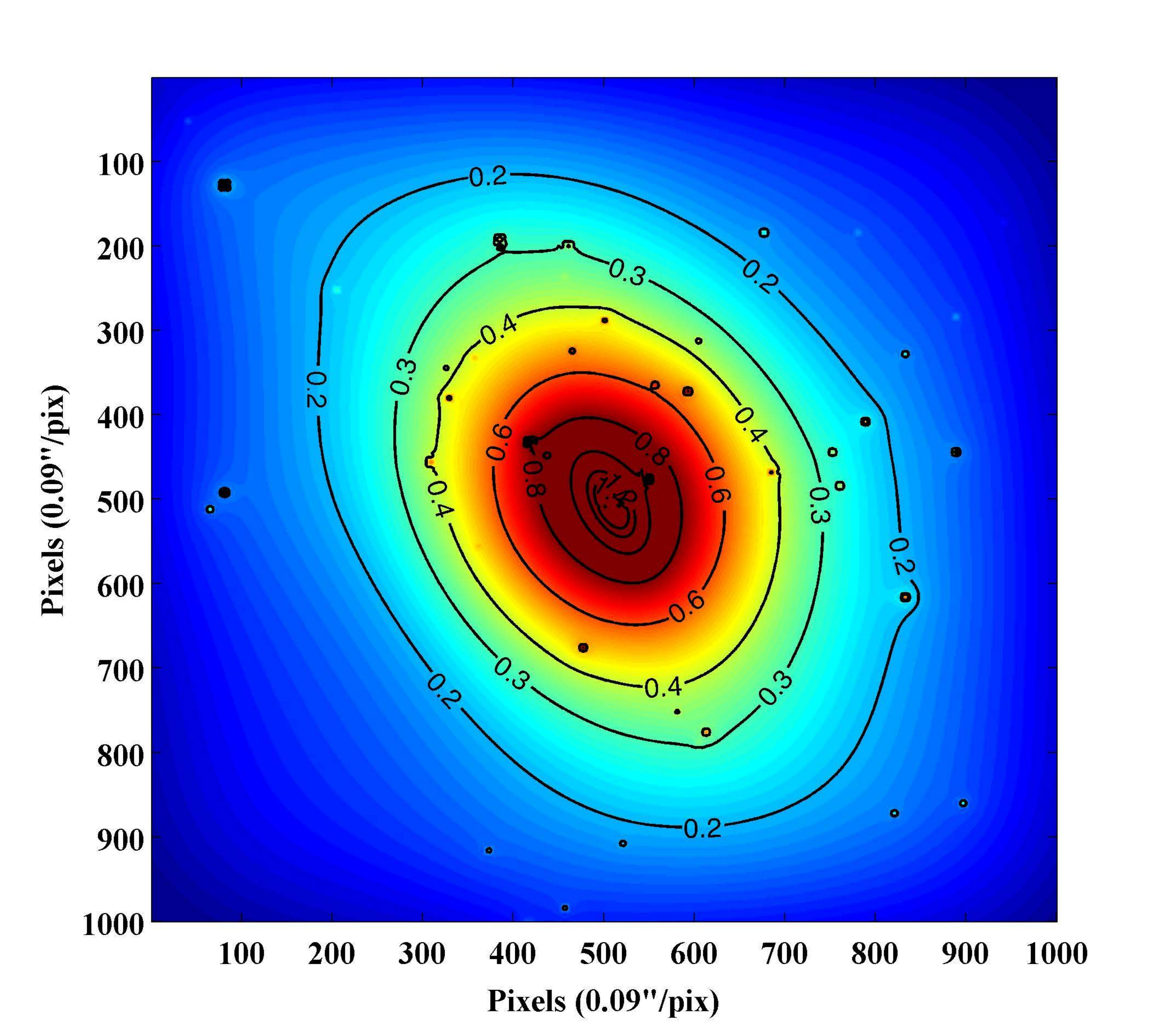}
    \includegraphics[width=82mm,trim=2cm 1cm 2cm 2cm,clip]{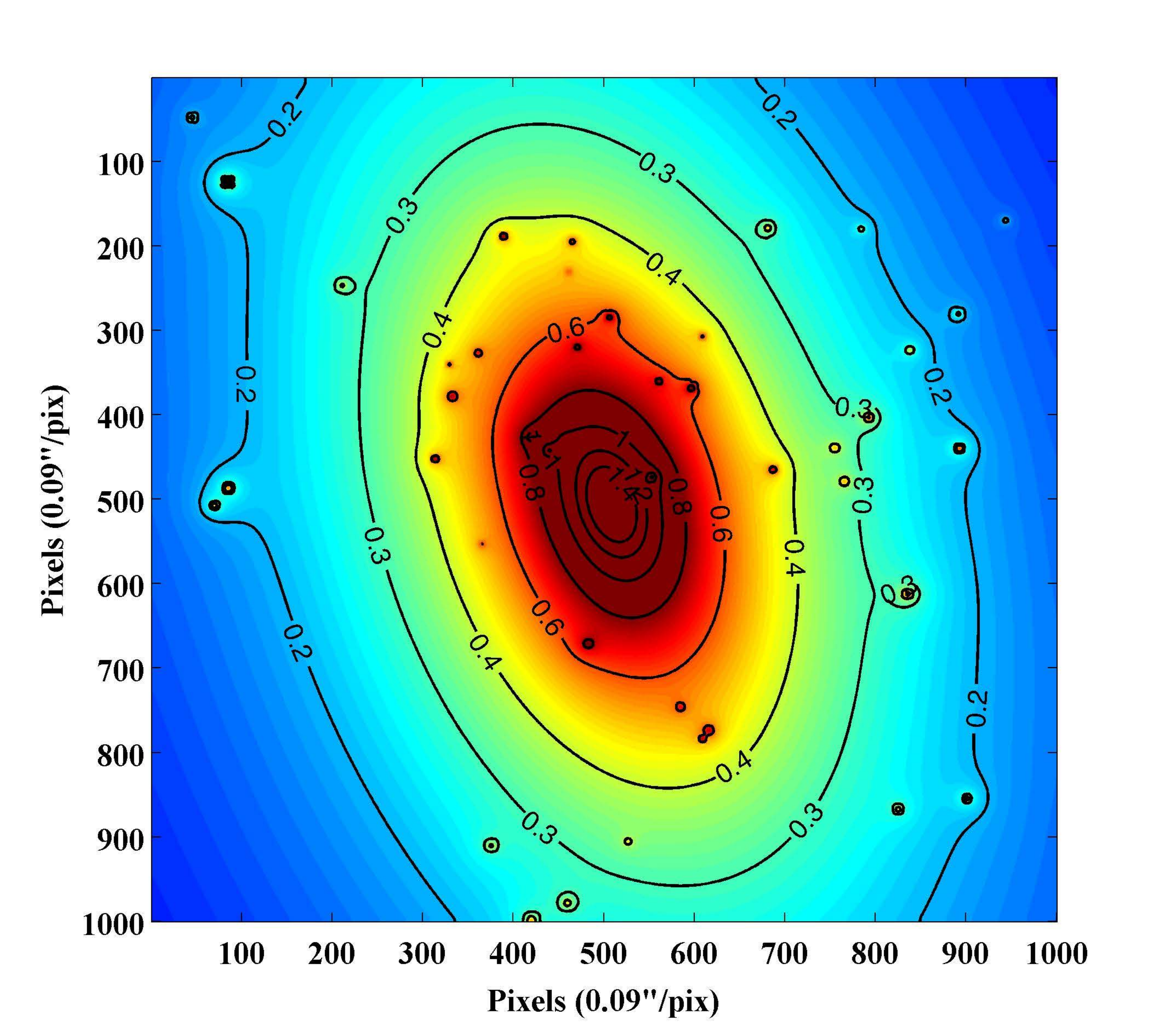}
       \includegraphics[width=82mm,trim=2cm 1cm 2cm 2cm,clip]{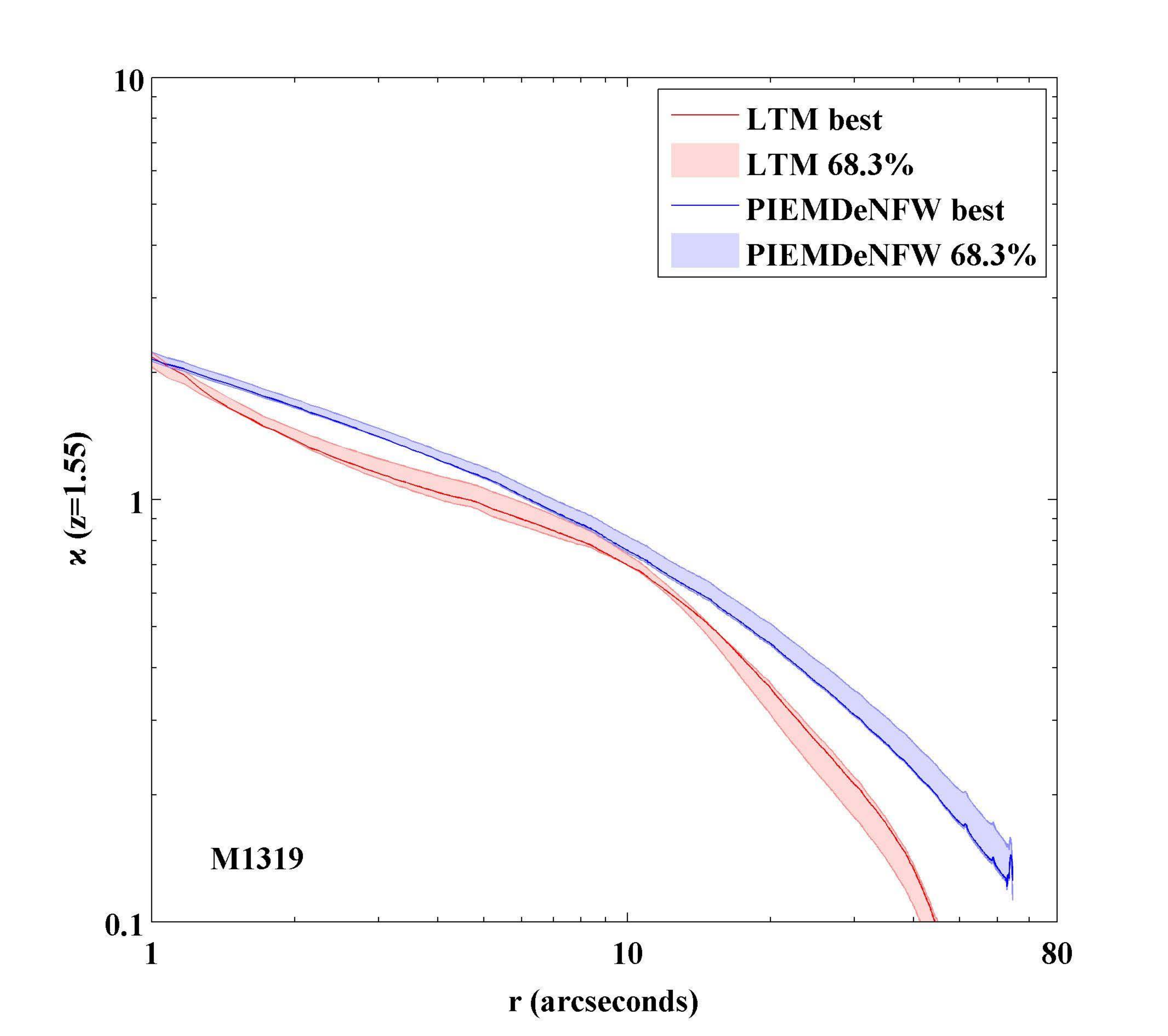}
\caption{Resulting mass models. Upper panel shows the mass-density \emph{kappa} map for a source at $z_{s}=1.55,$ the redshift of system 1, for the LTM model; the Middle Panel for the PIEMDeNFW model; and the Bottom Panel shows the resulting \emph{kappa} profile from the two models. Some notable differences are seen, which are, however, not surprising given the low number of constraints and different parametrizations. For further discussion on differences between the methods see \citet{Zitrin2014CLASH25,Meneghetti2016Comparison}.}
\label{fig3}
\end{figure}

The two mass distributions and profiles (Fig. \ref{fig3}) are in rough agreement -- with some differences expected given their different parametrizations and the small number of constraints available. In that sense they can be referred to as preliminary models. Both models however agree well -- to within 5\% -- regarding the size of the lens (see Fig. \ref{fig1}): we measure an effective Einstein radius of $\theta_{e}(z=1.55)\simeq11\arcsec$ for the redshift of system 1, and $\theta_{e}(z=3.52)=14\arcsec$ for that of system 2. The critical curves for these redshifts enclose $\simeq1.8\times10^{13}$ and $\simeq2.6\times10^{13}$ $M_{\odot}$, respectively, and the two models agree within 10\% on these mass measurements. For $z_{s}=2$, a value often used for comparison, we find $\theta_{e}(z=2)\simeq12\arcsec$ enclosing $\simeq2.1\times10^{13}$ $M_{\odot}$. Note the nominal uncertainties we typically adopt for these quantities are 10\% on the Einstein radii and 15\% on the enclosed mass. These nominal uncertainties are only slightly higher than the typical statistical uncertainties but encompass better the underlying systematics \citep{Zitrin2014CLASH25}.

Note that the final \emph{rms} of our pipeline is often somewhat higher than in other schemes: the LTM model, and for self-consistency purposes also the fully-parametric PIEMDeNFW model, are in practice constructed on a grid, whose resolution is, for speed-up purposes, comparable to or somewhat lower than that of HST. In significant magnification regions the round-up of the average source position to the grid's lower resolution pixel scale, introduces a finite, non-negligible \emph{rms} error of order 0.1\arcsec\ per system, contributing quite significantly to the global, quoted imprecision of the model (but, importantly, without harming its reliability nor prediction power). These points have been recently emphasized in more length in a community effort to compare lens modeling techniques to simulated clusters \citep{Meneghetti2016Comparison}, and we refer the interested reader to that work for more discussion on this end\footnote{Also note we aim to improve this numerically in the near future.}.

\vspace{0.05cm}

\vspace{0.3cm}
\section{Discussion and Summary} \label{discussion}

M1319 is a massive galaxy cluster, with an X-ray inferred mass of $M_{500}=4.8\pm0.9\times10^{14}$ M$_{\odot}$ \citep{Mantz2010}, and a measured velocity dispersion of $\sim$1000 km/s, in good agreement with its weak-lensing (WL) measurement suggesting $\sigma_{WL}=1160\pm140$ km/s \citep{Irgens2002WL}. Naturally, not all massive clusters have SL regions in proportion to their overall mass. For maximizing the SL properties there is great importance to how the matter is distributed within the cluster, for example, its concentration \citep{Broadhurst2008} and elongation along the line of sight \citep[e.g.][]{Hennawi2007,Sereno2010,Merten2015CLASH}, or alternatively, if there are substantial mass clumps and/or effective ellipticity boosting the critical area and lensing cross section \citep[e.g.][]{Redlich2012MergerRE,Zitrin2013M0416}.

Part of the motivation for our work here is to systematically map cluster lenses with archival HST imaging, so that the best cosmic telescopes could be designated before the launch of JWST. M1319 lies at high ecliptic latitude where the zodiacal IR background is low, which might be beneficial for JWST studies of high-redshift galaxies. For $z_{s}=15$, for example, we find an effective Einstein radius of $\theta_{e}(z=15)\simeq16\arcsec$, enclosing $3.1\times10^{13}$ $M_{\odot}$. This is a relatively small lens size compared to other massive clusters (MACS clusters in particular, e.g. \citealt{Zitrin2016MACS2135}, or those selected for the Hubble Frontier Fields program, see \citealt{Lotz2016HFF}), so that our analysis revealed M1319 is perhaps not in the top class of lensing clusters. Nonetheless, while larger lenses may be more efficient, also somewhat smaller lensing clusters such as M1319 are worth observing, and can usefully magnify faint background sources with -- in this case -- the advantage that most moderately-magnified region lies well within HST's (and JWST's) near-infrared CCDs. 

Another main motivation to studying this massive cluster followed the detection of the SG. We now estimate approximately the chances of seeing such a system forming in a galaxy cluster. To form such a symmetric well-aligned SG, the encounter should occur with an impact parameter of the scale of the host's core. This renders the cross section ($\sigma_{SG}=\pi r_{core}^2$) for such a configuration of order kpc$^{2}$ (adopting a galaxy core radius of $\sim0.5-1$ kpc).  The resulting mean free path before such an event, $l=1/(n\sigma_{SG})$, where $n$ is the number density of galaxies which we take as $\sim1000$ Mpc$^{-3}$, a typical thumb number for massive clusters, comes out to be of order $l\sim1/(10^{-2})\sim1000$ Mpc. In contrast, the radius of massive clusters is of order $\sim1$ Mpc, including that of M1319 \citep{Mantz2010}, which means each galaxy has order tenth of a percent to become a SG, in each crossing of the cluster (in general the crossing time is of order Gyr so that only few crossings per galaxy are expected). To obtain the chances a cluster would produce a SG we need to multiply by the number of galaxies in the cluster for which we take a nominal 1000 galaxies per cluster. However, we only need to account for the fraction of galaxy pairs with low enough relative velocities. We assume a (radial) velocity dispersion of 1000 km/s and account only for velocities -- with respect to the mean velocity -- lower than the escape velocity from the SG host which we take as 200 km/s. This yields to first order approximation $(200/1000)^3\sim1\%$ of the galaxies (or an order of magnitude less, if actually counting only the possible pairs rather then approximating as above). Assuming the dissipation time scale of the shells, i.e. the timeframe in which the shells can be observed after having formed, is of order Gyr, in total we get that the chances to see a shell galaxy is of order one in a few dozen to one in a few hundred massive clusters. Note that we neglected the mass distribution of cluster galaxies and did not demand certain mass ratios. Note also that the estimate is susceptible to the different assumptions, especially the galaxy core radius (affecting the cross section per galaxy) or escape velocity, that within a reasonable value range can easily change the estimate by an order-of-magnitude. Overall, this calculation shows why SGs are rare in clusters (note however that we do not refer to the BCG in our estimate here, for which other assumptions may apply). A search for SGs in archival HST imaging of other massive clusters would be interesting, to confront and reassess this estimate.

\begin{figure*}
 \begin{center}
   \includegraphics[width=177mm,trim=0cm 0cm 0cm 0cm,clip]{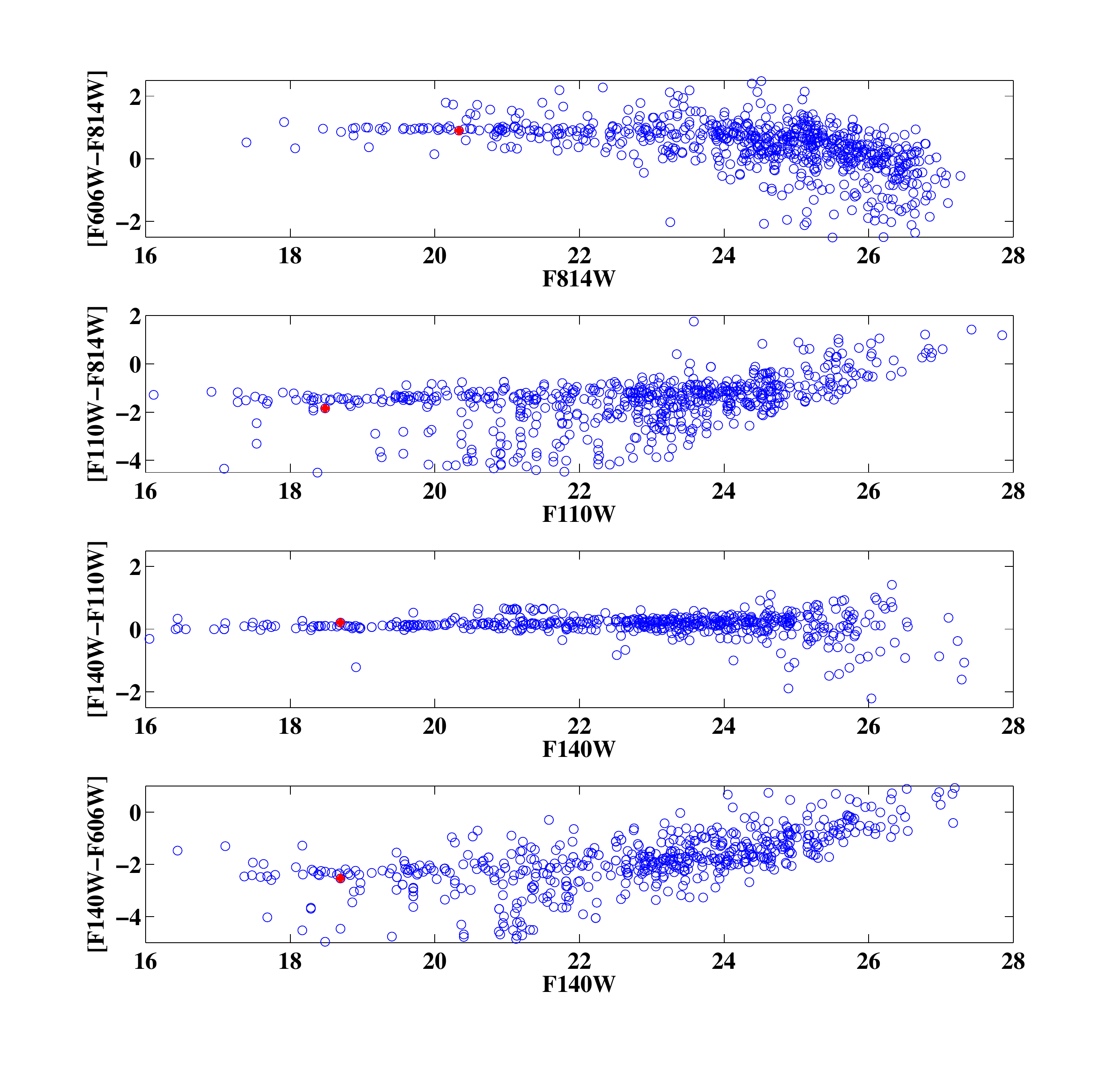}
 \end{center}
\caption{The Shell Galaxy in Color-Magnitude space. Figure shows four different color-magnitude diagrams from photometric catalogs generated for the central field of M1319. We plot all objects (\emph{blue open circles}) cross-matched between the different bands in the central 1.5'$\times$1.5' field. The SG, marked with \emph{filled red}, lies exactly on the top of easily-identifiable cluster-member red sequence, leaving little doubt it is indeed a cluster member. Future spectroscopic redshifts will help to confirm his assumption.}\vspace{0.2cm}
\label{fig4}
\end{figure*}

We can exploit our mass model's best-fit M/L scaling relation for cluster galaxies to estimate the masses of the SG. Our mass model suggests a total mass of $\sim1.3\times10^{11}$ M$_{\odot}$ for the SG system, yielding a mass-to-light ratio of $M/L_{B}\sim15$; a typical value for cluster galaxies. The luminosities, or magnitudes, of the SG host (F814W$=$20.61 AB) and companion (F814W$=$22.50 AB) suggest a minor merger of mass ratio of roughly 10:1. Clearly, this is an upper limit as some stars of the companion are already distributed to the shells, so it has been somewhat more massive to begin with, than its current luminosity suggests. While we leave detailed modeling of this system to future work, from the mere fact that both the host and progenitor are still observed, and that only two shells are seen, one of them half the distance of the other from the center, it is immediately implied that this a relatively young system compared to the expected merger timescale for this mass ratio (typically of order several Gyr, see for example \citealt{Ebrova2013Thesis,Lotz2008timescale,Boylan-Kolchin2008,Jiang2008}). Indeed, new generations of numerical simulations are now capable of simulating complex SG systems with high resolution \citep{Cooper2011ShellSimNGC7600Like,Ebrova2012LineProfilesShells}. Given the rarity of SGs in massive clusters, and perhaps accompanied by our public mass model, it might be interesting to dedicatedly simulate this system in future work.


\section*{acknowledgments}
AZ thanks the reviewer of this work for useful comments. AZ thanks Sirio Belli for his contribution to the Keck/MOSFIRE observations and analysis. AZ is very grateful for a proof read of this manuscript and insightful comments by Ivana Ebrova. While we defer numerically simulating the SG in detail to future work, AZ is indebted to Chris Hayward for help in simulations setup, and acknowledges useful discussions with Margaret Geller, Richard Ellis, Sterl Phinney, Andrew Wetzel, Cameron Hummels, Phil Hopkins, Benny Trakhtenbrot, Dan Stern, Tom Broadhurst, Holland Ford, Re'em Sari and Harald Ebeling. Comments received from Michal Bilek are appreciated. Principal support for this work was provided by NASA through Hubble Fellowship grant \#HST-HF2-51334.001-A awarded by STScI, which is operated by the Association of Universities for Research in Astronomy, Inc. under NASA contract NAS~5-26555. This work is in part based on previous observations made with the NASA/ESA Hubble Space Telescope. Data presented herein were obtained at the W.M. Keck Observatory. The authors wish to recognize and acknowledge the very significant cultural role and reverence that the summit of Mauna Kea has always had within the indigenous Hawaiian community. We are most fortunate to have the opportunity to conduct observations from this mountain.


\end{document}